# Fractional Generalization of Higher-Order Diffusion

K. Parisis[1] and E. C. Aifantis [1-3]

[1]Aristotle University of Thessaloniki, Thessaloniki, 54124, Greece,
[2]Michigan Technological University, Houghton, MI 49931, USA
[3]Togliatti State University, Togliatti 445020, Russia

*mom@mom.gen.auth.gr, ORCID: 0000-0002-6846-5686, Tel.:+30-2310995921*

## Abstract

A fractional generalization of the second author's higher-order diffusion theory is given and fundamental solutions are obtained. The extension from the integer to the fractional case involves a proper treatment of the fractional Laplacian of Riesz type. This is done with analogy to an earlier treatment in extending the second author's gradient elasticity model from the integer to the fractional case.

## 1. Introduction

A fractional generalization of higher-order diffusion theory is provided. Such theory was proposed in the 1980's by the second author [1] on a purely mechanical basis using Maxwell's concept of diffusive force (introduced in his statistical treatment of polyatomic gases) and the usual statement of momentum balance for each species separately. It ends up with various classes of mass transport models generalizing the classical Fick's law, including Barenblatt's pseudoparabolic equation of fluid flow in fissured rocks, Cateano's equation of heat conduction, and the Cahn-Hilliard equation of spinodal decomposition. It was used, among other things, to model grain boundary diffusion and mass transport in nanopolycrystals [2]. An alternative formulation, along more conventional lines, for deriving higher-order diffusion equations is to generalize Fick's classical law for the diffusive flux by introducing additional and physically motivated gradients of the density (concentration) of the diffusing species. For the specific model of higher-order diffusion to be dealt with herein, this amounts to supplementing the mass balance equation

$$\frac{\partial \rho}{\partial t} + div\,\mathbf{j} = 0 \quad or \quad \frac{\partial \rho}{\partial t} + j_{i,i} = 0\,, \tag{1.1}$$

for the density $\rho$, and the flux $\mathbf{j}$ of the diffusive substance, with the following higher-order constitutive equation

$$\mathbf{j} = -D\nabla(\rho - l_d^2 \nabla^2 \rho);\ j_i = -D\,(\rho - l_d^2 \nabla^2 \rho)_{,i}\,, \tag{1.2}$$

where $D$ is the classical diffusion coefficient and the new parameter $l_d$ denotes an internal length accounting for "weakly" nonlocal diffusion effects.

The fractional generalization of the above equation consists of replacing the standard (integer) Laplacian $\Delta \equiv \nabla^2$ in Eq. (1.2) with a fractional one of the Riesz form $(-^R\Delta)^{\alpha/2}$ or the Caputo form $^C\Delta_W^\alpha$. Then, the corresponding fractional generalization of Eq. (1.2) reads

$$\mathbf{j} = -D\nabla[\rho - l_d^2(\alpha)\{(-^R\Delta)^{\alpha/2}\rho\}];\ j_i = -D\,[\rho - l_d^2(\alpha)\{(-^R\Delta)^{\alpha/2}\rho\}]_{,i}\,, \tag{1.3}$$

and

$$\mathbf{j} = -D\nabla[\rho - l_d^2(\alpha)\{(^C\Delta_W^\alpha)\rho\}];\ j_i = -D[\rho - l_d^2(\alpha)\{(^C\Delta_W^\alpha)\rho\}]_{,i}\,, \tag{1.4}$$

respectively.

On introducing the aforementioned fractional gradient constitutive equations into the non-fractional balance law given by Eq. (1.1), corresponding partial differnetial equations of fractional order are obtained which need to be solved with the aid of appropriate boundary conditons for finite domains. To dispense with the complication of higher-order fractional boundary conditions, we consider infinite domains and derive fundamental solutions for the respective problems by employing a fractional extension of the Green's function method.

## 2. Fractional Second-Order Diffusion Equation

On introducing the non-classical fractional diffusion constitutive relation given by Eq. (1.3) into the classical (non-fractional) mass balance law, given by Eq. (1.1), we obtain the fractional high-order diffusion equation

$$\frac{\partial \rho}{\partial t} = D \Delta \rho + D l_d^\alpha \nabla \cdot \{(-\Delta)^{\alpha/2} \nabla \rho\}, \tag{2.1}$$

along with the auxiliary conditions $\rho(\mathbf{r},0) = \delta(\mathbf{r}), \rho(\mathbf{r},t) \to 0$ as $|\mathbf{r}| \to \infty$, where $\delta(\mathbf{r})$ denotes the delta function and the notation $[l_d^2(\alpha) \equiv l_d^\alpha, (-{}^R\Delta)^{\alpha/2} \equiv (-\Delta)^{\alpha/2}]$ was used for simplicity.

To solve Eq. (2.1) we employ the method of Fourier transform and exploit the properties of the Riesz fractional Laplacian, along with the symmetry of the problem. This gives

$$\frac{\partial \rho(\mathbf{k},t)}{\partial t} = -D|\mathbf{k}|^2 \rho(\mathbf{k},t) - D l_d^\alpha |\mathbf{k}|^2 |\mathbf{k}|^\alpha \rho(\mathbf{k},t), \tag{2.2}$$

where $\mathbf{k}$ denotes the wave vector. Equation (2.2) is a first order ordinary differential equation with respect to time with initial condition $\rho(\mathbf{k},0) = \mathcal{F}\{\delta(\mathbf{r})\}(\mathbf{k}) = 1$, with $\mathcal{F}$ denoting the Fourier transform. Its solution is

$$\frac{\partial \rho(\mathbf{k},t)}{\partial t} = \exp(-Dt|\mathbf{k}|^2) \times \exp(-D_\alpha t |\mathbf{k}|^{\alpha+2}), \tag{2.3}$$

where we defined $D_\alpha \equiv D l_d^\alpha$. The solution of Eq. (2.3) in configuration space is obtained by inversion of the Fourier transform

$$\rho(\mathbf{r},t) = \frac{1}{(2\pi)^3} \int_{-\infty}^{\infty} \exp(-Dt|\mathbf{k}|^2) \exp(-D_\alpha t |\mathbf{k}|^{\alpha+2}) \exp(i\mathbf{k}\cdot\mathbf{r}) d^3\mathbf{k}. \tag{2.4}$$

Equation (2.4) is the inverse Fourier transform of the product of two independent terms and can be expressed as the convolution of the corresponding solutions in the physical space using the following well-known property of the Fourier transform

$$\mathcal{F}\{(f*g)(\mathbf{r},t)\}(\mathbf{k},t) = \mathcal{F}\{f(\mathbf{r},t)\}(\mathbf{k},t) \mathcal{F}\{g(\mathbf{r},t)\}(\mathbf{k},t) \tag{2.5}$$

$$(f*g)(\mathbf{r},t) = \int_{-\infty}^{\infty} f(\mathbf{r}-\mathbf{r}',t) g(\mathbf{r}',t) d^3\mathbf{r}', \tag{2.6}$$

where the symbol $*$ denotes the usual binary operator of convolution of two integrable functions, and $\mathcal{F}$ denotes the Fourier transform. Using Eq. (2.5), we recognize Eq. (2.4) as the convolution

$$\rho(\mathbf{r},t) = (G_2 * G_{\alpha+2})(\mathbf{r},t), \tag{2.7}$$

where we defined the set of functions $G_\alpha(\mathbf{r},t)$ as

$$G_\alpha(\mathbf{r},t) = \mathcal{F}^{-1}\{\exp(-D_\alpha t |\mathbf{k}|^\alpha)\}$$
$$= \frac{1}{(2\pi)^3} \int_{-\infty}^{\infty} \exp(-D_\alpha t |\mathbf{k}|^\alpha) \exp(i\mathbf{k}\cdot\mathbf{r}) d^3\mathbf{k}, \qquad (2.8)$$

Equation (2.8) is the fundamental solution (i.e. the Green's function) for the fractional diffusion equation

$$\frac{\partial G_\alpha(\mathbf{r},t)}{\partial t} = -D_\alpha (-\Delta)^{\alpha/2} G_\alpha(\mathbf{r},t). \qquad (2.9)$$

The corresponding fundamental solution of Eq. (2.1) is then deduced from Eq. (2.8) through the convolution of Eq. (2.7).

The integral given by Eq. (2.8) is defined in a 3-dimensional Euclidean space and can be analytically computed in spherical coordinates by applying the well-known relationship (see, for example, Lemma 25.1 of Samko et al [3])

$$\frac{1}{(2\pi)^3} \int f(|\mathbf{r}|) e^{i\mathbf{k}\cdot\mathbf{r}} d^3\mathbf{k} = \frac{1}{(2\pi)^{3/2} |\mathbf{r}|^{1/2}} \int_0^\infty k^{3/2} f(k) J_{1/2}(k|\mathbf{r}|) dk. \qquad (2.10)$$

In Eq.(2.10) $k$ denotes the magnitude of the wave vector and $J_{1/2}(z) = \sqrt{2/(\pi z)} \sin(z)$ denotes the Bessel function of order ½. Applying Eq. (2.10) into the fundamental solution $G_\alpha(\mathbf{r},t)$ of Eq. (2.8), we obtain

$$G_\alpha(\mathbf{r},t) = \frac{1}{(2\pi)^{3/2} |\mathbf{r}|^{1/2}} \int_0^\infty k^{3/2} \exp(-D_\alpha t k^\alpha) J_{1/2}(k|\mathbf{r}|) dk,$$
$$= \frac{1}{2\pi^2 |\mathbf{r}|} \int_0^\infty k \exp(-D_\alpha t k^\alpha) \sin(k|\mathbf{r}|) dk. \qquad (2.11)$$

The integral in Eq. (2.11) can be computed using the convolution property of the Mellin transform, defined in [4] by the relationship

$$M\{f(x)\}(s) = \int_0^\infty f(x) x^{s-1} dx. \qquad (2.12)$$

Its inverse is given by

$$f(x) = \frac{1}{2\pi i} \int_{\gamma-i\infty}^{\gamma+i\infty} f(s) x^{-s} ds, \qquad (2.13)$$

where the path of integration is a vertical strip separating the poles of $M\{f(x)\}(s)$, defined in $\gamma_1 < \text{Re}(s) < \gamma_2$. For more details about the Mellin transform, we refer the reader to [5]. Here we only use the basic results

$$M\{\exp(-D_\alpha t k^\alpha)\}(s) = (D_\alpha t)^{-s/\alpha} \frac{1}{\alpha} \Gamma(\frac{s}{\alpha}),$$

$$M\{x^{3/2} J_{1/2}(x)\}(s) = 2^{s+1/2} \Gamma(1+\frac{s}{2}) / \Gamma(\frac{1}{2}-\frac{s}{2}), \tag{2.14}$$

where we made use of the Mellin transform of the Bessel function (see also Section 6.8 of [4])

$$M\{J_\nu(2\sqrt{x})\}(s) = \Gamma(\frac{\nu}{2}+s) / \Gamma(\frac{\nu}{2}+1-s). \tag{2.15}$$

Consequently, Eq. (2.11) can be evaluated using the above results and performing the inverse Mellin transform by computing the Mellin-Barnes integral

$$G_\alpha(\mathbf{r},t) = \frac{(D_\alpha t)^{-1/\alpha}}{2\alpha \pi^{3/2} |\mathbf{r}|^2} \frac{1}{2\pi i} \int_{\gamma-i\infty}^{\gamma+i\infty} \frac{\Gamma(\frac{1}{\alpha}-\frac{s}{\alpha})\Gamma(1+\frac{s}{2})}{\Gamma(\frac{1}{2}-\frac{s}{2})} (\frac{|\mathbf{r}|}{2(D_\alpha t)^{1/\alpha}})^{-s} ds. \tag{2.16}$$

The Mellin-Barnes integral representation of Eq. (2.16) can be expressed in terms of the corresponding Fox-H function of fractional analysis (see, for example, [6-9])

$$G_\alpha(\mathbf{r},t) = \frac{(D_\alpha t)^{-1/\alpha}}{2\alpha \pi^{3/2} |\mathbf{r}|^2} H^{1,1}_{1,2}\left[\frac{|\mathbf{r}|}{2(D_\alpha t)^{1/\alpha}}; \begin{array}{c}(1-\frac{1}{\alpha},\frac{1}{\alpha})\\(1,\frac{1}{2}),(\frac{1}{2},\frac{1}{2})\end{array}\right]. \tag{2.17}$$

After a change of variables $s \to s-2$, Eq. (2.16) becomes

$$G_\alpha(\mathbf{r},t) = \frac{1}{\alpha (4\pi)^{3/2} (D_\alpha t)^{3/\alpha}} \frac{1}{2\pi i} \int_{\gamma-i\infty}^{\gamma+i\infty} \frac{\Gamma(\frac{3}{\alpha}-\frac{s}{\alpha})\Gamma(\frac{s}{2})}{\Gamma(\frac{3}{2}-\frac{s}{2})} (\frac{|\mathbf{r}|}{2(D_\alpha t)^{1/\alpha}})^{-s} ds. \tag{2.18}$$

By using the substitution $s \to 2s$, Eq. (2.18) can be rewritten as

$$G_\alpha(\mathbf{r}) = \frac{2}{\alpha (4\pi)^{3/2} (D_\alpha t)^{3/\alpha}} \frac{1}{2\pi i} \int_{\gamma-i\infty}^{\gamma+i\infty} \frac{\Gamma(\frac{3}{\alpha}-\frac{2s}{\alpha})\Gamma(s)}{\Gamma(\frac{3}{2}-s)} (\frac{|\mathbf{r}|}{2(D_\alpha t)^{1/\alpha}})^{-2s} ds. \tag{2.19}$$

The Mellin-Barnes integral of Eq. (2.19) can be evaluated by applying the residue theorem, following the standard procedure [8]. The final result is the following series expansion expression

$$G_\alpha(\mathbf{r},t) = \frac{2}{\alpha(4\pi)^{3/2}(D_\alpha t)^{3/\alpha}} \sum_{\nu=0}^{\infty} \frac{(-1)^\nu}{\nu!} \frac{\Gamma(\frac{3}{\alpha}+\frac{2\nu}{\alpha})}{\Gamma(\frac{3}{2}+\nu)} \left(\frac{|\mathbf{r}|^2}{4(D_\alpha t)^{2/\alpha}}\right)^\nu. \qquad (2.20)$$

Equation (2.20) can be represented in terms of the Wright's function $_1\Psi_1$ as

$$G_\alpha(\mathbf{r},t) = \frac{2}{\alpha(4\pi)^{3/2}(D_\alpha t)^{3/\alpha}} \,_1\Psi_1\left[\begin{array}{c}(\frac{3}{\alpha},\frac{2}{\alpha})\\(\frac{3}{2},1)\end{array}; -\frac{|\mathbf{r}|^2}{4(D_\alpha t)^{2/\alpha}}\right], \qquad (2.21)$$

where the generalized Wright's function $_p\Psi_q$ is defined by the following series [8-9]

$$_p\Psi_q(z) = \,_p\Psi_q\left[\begin{array}{ccc}(a_1,A_1) & \cdots & (a_p,A_p)\\(b_1,B_1) & \cdots & (b_q,B_q)\end{array}; z\right] = \sum_{\nu=0}^{\infty} \frac{\prod_{j=1}^{p}\Gamma(a_j+A_j\nu)}{\prod_{j=1}^{q}\Gamma(b_j+B_j\nu)} \frac{z^\nu}{\nu!}. \qquad (2.22)$$

It is easy to check that when $\alpha=2$, the series expansion reduces to the Green's function of the ordinary diffusion equation in 3-dimensional space. This is readily seen by letting $\alpha \to 2$ in Eq. (2.21), resulting to the expression

$$\begin{aligned}G_2(\mathbf{r},t) &= \frac{1}{(4\pi Dt)^{3/2}} \,_1\Psi_1\left[\begin{array}{c}(\frac{3}{2},1)\\(\frac{3}{2},1)\end{array}; -\frac{|\mathbf{r}|^2}{4Dt}\right]\\ &= \frac{1}{(4\pi Dt)^{3/2}} \sum_{\nu=0}^{\infty} \frac{(-1)^\nu}{\nu!} \left(\frac{|\mathbf{r}|^2}{4Dt}\right)^\nu \\ &= \frac{1}{(4\pi Dt)^{3/2}} \exp\left(-\frac{|\mathbf{r}|^2}{4Dt}\right).\end{aligned} \qquad (2.23)$$

Consequently, the fundamental solution of the second-order fractional diffusion equation (2.1), denoted as $G(\mathbf{r},t)$, is obtained through convolution of Eq. (2.7), with $G_\alpha(\mathbf{r},t)$ given by Eq. (2.21) and (2.23) for $\alpha=2$, i.e.

$$G(\mathbf{r},t) = \int_{-\infty}^{\infty} G_{\alpha+2}(\mathbf{r}-\mathbf{r}',t) G_2(\mathbf{r}',t) \, d^3\mathbf{r}'. \tag{2.24}$$

We can extend Eq. (1.1) to include distributed sources (e.g. chemical reaction or trapping) with density/concentration rate $q(\mathbf{r},t)$. In this particular case, the classical mass balance law becomes

$$\frac{\partial \rho}{\partial t} + div\, \mathbf{j} = q, \tag{2.25}$$

and the corresponding inhomogeneous fractional diffusion equation reads

$$\frac{\partial \rho}{\partial t} = D\Delta\rho + D l_d^\alpha \nabla \cdot \{(-\Delta)^{\alpha/2} \nabla\rho\} + q. \tag{2.26}$$

Using the Fourier transform method, we can obtain the fundamental solution of Eq. (2.26) as follows

$$\rho(\mathbf{r},t) = \int_0^t \int_{-\infty}^{\infty} G(\mathbf{r}-\mathbf{r}',t-\tau) q(\mathbf{r}',\tau) \, d^3\mathbf{r}' d\tau, \tag{2.27}$$

where $G(\mathbf{r},t)$ is given by Eq. (2.24). For the special case of a unit point source $q(\mathbf{r},t) = \delta(\mathbf{r})\delta(t)$, Eq. (2.27) reduces to the fundamental solution $G(\mathbf{r},t)$.

The fractional diffusion equation admits steady-state solutions, under the presence of external source/sink terms with density/rate $q(\mathbf{r})$. The governing equation for this time independent configuration is

$$D\Delta\rho + D l_d^\alpha \nabla \cdot \{(-\Delta)^{\alpha/2} \nabla\rho\} + q(\mathbf{r}) = 0. \tag{2.28}$$

Equation (2.28) can be generalized to a higher-order steady-state fractional diffusion equation of the form

$$D_\alpha ((-\Delta)^{\alpha/2} \rho)(\mathbf{r}) + D_\beta ((-\Delta)^{\beta/2} \rho)(\mathbf{r}) = q(\mathbf{r}); \quad (\alpha > \beta), \tag{2.29}$$

where $(\alpha,\beta)$ denote arbitrary positive fractional order and $(D_\alpha, D_\beta)$ are corresponding fractional diffusion coefficients. Equation (2.29) can be derived by considering a fractional extension of the conservation law given by Eq. (2.25), along with the consitutive relation given by Eq. (1.3) and/or a further fractional extension for its classical gradient ($\nabla$) part.

Equation (2.29) is a fractional partial differential equation, whose solution reads

$$\rho(\mathbf{r}) = \int G_{\alpha,\beta}(\mathbf{r}\text{-}\mathbf{r}')q(\mathbf{r}')d^3\mathbf{r}', \qquad (2.30)$$

with the Green-type function $G_{\alpha,\beta}(r)$ given by

$$G_{\alpha,\beta}(r) = \frac{1}{(2\pi)^3}\int \frac{1}{D_\alpha |k|^\alpha + D_\beta |k|^\beta} e^{i k \cdot r} d^3 k = \frac{1}{(2\pi)^{3/2}\sqrt{|r|}}\int_0^\infty \frac{\lambda^{3/2} J_{1/2}(\lambda|r|)}{D_\alpha \lambda^\alpha + D_\beta \lambda^\beta}d\lambda. \qquad (2.31)$$

Let us now consider the particular problem of a unit point source located at the origin of the form

$$q(\mathbf{r}) = q_0 \delta(\mathbf{r}) = q_0 \delta(x)\delta(y)\delta(z). \qquad (2.32)$$

Upon substitution of Eq. (2.32) into Eq. (2.30) we obtain the particular solution

$$\rho(\mathbf{r}) = q_0 G_{\alpha,\beta}(\mathbf{r}), \qquad (2.33)$$

with the Green function $G_{\alpha,\beta}(r)$ given by Eq. (2.31). By using then the particular expression for the Bessel function of the first kind $J_{1/2}(z) = \sqrt{2/(\pi z)}\sin(z)$, we obtain

$$\rho(\mathbf{r}) = \frac{1}{2\pi^2}\frac{q_0}{|r|}\int_0^\infty \frac{\lambda \sin(\lambda|r|)}{D_\alpha \lambda^\alpha + D_\beta \lambda^\beta}d\lambda; \quad (\alpha > \beta). \qquad (2.34)$$

Two disctinct modes of diffusion arise, depending on the particular form of the fractional parameters $(\alpha,\beta)$, which are discussed in detail below:

*A) Sub- GradDiffusion:* $\alpha = 2; 0 < \beta < 2$. In this case Eq. (2.29) becomes

$$D\Delta\rho(\mathbf{r}) - D_\beta ((-\Delta)^{\beta/2}\rho)(\mathbf{r}) + q(\mathbf{r}) = 0, \quad (0 < \beta < 2). \qquad (2.35)$$

The order of the fractional Laplacian $(-\Delta)^{\beta/2}$ is less than the order of the first term related to the usual Fick's law. The parameter $\beta$ defines the order of the power-law non-locality. The particular solution of Eq. (2.35) in the present case, reads

$$\rho(\mathbf{r}) = \frac{1}{2\pi^2}\frac{q_0}{|r|}\int_0^\infty \frac{\lambda \sin(\lambda|r|)}{D\lambda^2 + D_\beta \lambda^\beta}d\lambda; \quad (0 < \beta < 2). \qquad (2.36)$$

The following asymptotic behavior for Eq. (2.36) can be derived in the form

$$\rho(\mathbf{r}) = \frac{q_0}{2\pi^2 |r|}\int_0^\infty \frac{\lambda \sin(\lambda|r|)}{D\lambda^2 + D_\beta \lambda^\beta}d\lambda \approx \frac{C_0(\beta)}{|r|^{3-\beta}} + \sum_{k=1}^\infty \frac{C_k(\beta)}{|r|^{(2-\beta)(k+1)+1}}, \quad (|r| \to \infty), \qquad (2.37)$$

where

$$C_0(\beta)=\frac{q_0}{2\pi^2 D_\beta}\Gamma(2-\beta)\sin\left(\frac{\pi}{2}\beta\right); C_k(\beta)=-\frac{q_0 D^k}{2\pi^2 D_\beta^{k+1}}\int_0^\infty z^{(2-\beta)(k+1)-1}\sin(z)dz. \quad (2.38)$$

As a result, the density of the diffusive species generated by the source that is concentrated at a single point in space, for large distances from the source, is given asymptotically by the expression

$$\rho(r)\approx\frac{C_0(\beta)}{|r|^{3-\beta}}; \quad (0<\beta<2). \quad (2.39)$$

*B) Super-GradDiffusion:* $\alpha>2$ and $\beta=2$. In this case, Eq. (2.29) becomes

$$D\Delta\rho(r)-D_\alpha((-\Delta)^{\alpha/2}\rho)(r)+q(r)=0; \quad (\alpha>2). \quad (2.40)$$

The order of the fractional Laplacian $(-\Delta)^{\alpha/2}$ is greater than the order of the first term related to the usual Fick's law. The asymptotic behavior of the density $\rho(r)$ for $|r|\to 0$ in this case is given by

$$\rho(r)\approx\begin{cases}\dfrac{q_0\Gamma((3-\alpha)/2)}{2^\alpha \pi^2 \sqrt{\pi} D_\alpha \Gamma(\alpha/2)}\dfrac{1}{|r|^{3-\alpha}}; & (2<\alpha<3),\\[2mm] \dfrac{q_0}{2\pi\alpha D^{1-3/\alpha} D_\alpha^{3/\alpha}\sin(3\pi/\alpha)}; & (\alpha>3).\end{cases} \quad (2.41)$$

Note that the above asymptotic behavior does not depend on the parameter $\beta$, and that the corresponding relation of Eq. (2.41) does not depend on $D_\beta$. The density $\rho(\mathbf{r})$ for short distances away from the point of source application is determined only by the term with $(-\Delta)^{\alpha/2}$ ($\alpha>2$), i.e. the fractional counterpart of the usual extra non-Fickean term of the fractional diffusion equation.

Finally, and especially for the case of more complicated boundary value problems, we mention that the steady-state diffusion of Eq. (2.28) can be factored as

$$D\nabla\cdot\nabla\{1+l_d^\alpha(-\Delta)^{\alpha/2}\}\rho+q(\mathbf{r})=0. \quad (2.42)$$

By defining the "classical" operator $L^0\equiv D\nabla\cdot\nabla=D\,div\,grad$, and similarly its "fractional gradient" counterpart $L^\alpha\equiv 1+l_d^\alpha(-\Delta)^{\alpha/2}$, we can prove that $L^\alpha\rho$ satisfies the classical steady-state Fickean condition. This is a direct consequence of the fact that the operators $L^0$ and $L^\alpha$ commute. This allows us to write

$$L^0\{1+ l_d^\alpha (-\Delta)^{\alpha/2} \}\rho + q(\mathbf{r}) = 0. \tag{2.43}$$

Therefore, we arrive at the following "operator-split" scheme

$$(1+ l_d^\alpha (-\Delta)^{\alpha/2})\rho = \rho^0; \quad D\nabla\cdot\nabla\rho^0 + q(\mathbf{r}) = 0. \tag{2.44}$$

This is the fractional counterpart of the Ru-Aifantis theorem [10], for the steady-state fractional higher-order diffusion equation.

In concluding, we point out that this development of extending the higher-order diffusion theory form the integer to the fractional order follows the lines of such type of extension pursued earlier for the theory of gradient elasticity/GradEla [11-14].

## Acknowledgment

Support of the Ministry of Education and Science of Russian Federation under grant no. 14.Z50.31.0039 is acknowledged.